\documentclass[twocolumn,showpacs,preprintnumbers,aps]{revtex4} 

\def\XXint#1#2#3{{\setbox0=\hbox{$#1{#2#3}{\int}$}
     \vcenter{\hbox{$#2#3$}}\kern-.5\wd0}}

\usepackage{epsfig}
\begin{document} 
\preprint{FAU-TP3-04/01}
\title{Chiral Symmetry in Light-Cone Field Theory}  
\author{ F. Lenz$^1$, K. Ohta$^2$, M. Thies$^1$, and K. Yazaki$^3$} 
\address{$^{1}$Institut f\"ur Theoretische Physik III,  
Universit\"at Erlangen-N\"urnberg, 
D-91058 Erlangen, 
Germany \\
$^{2}$Institute of Physics, University of Tokyo, Komaba, Tokyo 153, Japan
\\ 
$^{3}$College of Arts and Sciences,
Tokyo Woman's Christian University,
Tokyo 167-8585, Japan
} 
\date{\today} 
\begin{abstract}
An analysis of spontaneously broken chiral symmetry in light-cone field theory is presented.
 The non-locality inherent to light-cone field theory requires revision of the standard procedure 
in the derivation of Ward-Takahashi identities. We derive  the general structure of chiral
 Ward-Takahashi identities and  construct them explicitly  for  various model field theories. 
Gell-Mann--Oakes--Renner relations and relations between fermion propagators and the structure
 functions of Nambu-Goldstone bosons are discussed and the necessary modifications of the
 Ward-Takahashi identities due to the axial anomaly are indicated.  
\end{abstract} 
\maketitle 
 


\section{Introduction}
Analysis of the  symmetry properties constitutes the most powerful tool in the  
study of physical systems. Symmetry properties constrain the structure of the theoretical
 description and have  far reaching consequences for the spectrum of states of the system. 
In systems with infinitely many degrees of freedom, symmetries can be realized in different
 ways. Very often, the different phases of physical systems are characterized by their
 realization of a symmetry. Of great importance is the case in which a continuous symmetry
 is not realized by the ground state of the system, {\em i.e.}, if the symmetry is spontaneously broken.
 The Goldstone theorem asserts that in this case the spectrum exhibits massless excitations. 
It thereby connects properties of the ground state of the system with the existence of
massless Nambu-Goldstone (NG) particles \cite{Nambu60,Goldstone61}. In the light-cone formulation of a field theory
 exhibiting the phenomenon of spontaneous symmetry breakdown  the connection between
 ground state properties and the spectrum of excitations appears to be lost 
(some early studies within the parton model can be found in \cite{Cash71,Cash73,Cash74}; 
for a review of light-cone quantization, see \cite{BPP98}). The ground
 state is determined kinematically, its properties are independent of the dynamics, in particular
 they are independent of the realization of the symmetry. Despite this deficiency in characterizing
 the system by the structure of the ground state, light-cone formulations of model field theories
 like the chiral Gross-Neveu \cite{GN74}, the 't~Hooft \cite{tHooft74} or the Nambu--Jona-Lasinio \cite{NJL61}
 model exhibit massless
 particles. Similarly, if string theory is solved in light-cone quantization,  the spectrum of
 excitations is correctly predicted. Also the successful application of light-cone field theory
 in the analysis of deep inelastic scattering on strongly interacting systems which are built
 upon the non-trivial QCD ground state indicates compatibility of the kinematical light-cone
 vacuum with non-trivial dynamics.

In this work we study  symmetry properties of systems with  spontaneously broken chiral
 symmetry. As in standard field theory, we will formulate the symmetry properties as chiral 
Ward-Takahashi (WT) identities \cite{Ward50,Takahashi57}. Such an analysis is most conveniently carried out with the 
help of functional techniques. In this framework we define a fermionic light-cone  field theory 
by the generating functional containing only the unconstrained degrees of freedom, {\em i.e.}, with
 the ``bad'' components of the spinors eliminated. Such a fermionic theory is  non-local,
 irrespective of the dynamics. This non-locality reflects the  dependence of the light-cone 
energy $p_+$  on the  inverse of the light-cone momentum $p_-$ in the dispersion relation
 of a free particle
$$p_+ = \frac{m^2 + p_{\perp}^2}{2p_-}\, .$$     
Kinematical nature of the vacuum and  non-locality of light-cone field theory have the 
same origin.  

In the standard derivation of the relevant WT identities one considers soft space-time 
dependent modulations of the global symmetry transformations \cite{Miransky93}.  In the symmetry broken 
phase, these transformations generate excitations with excitation energies approaching 
zero with increasing wavelength of the modulations. Via the WT identities one identifies 
 the creation operator of the NG bosons in the limit of vanishing momentum
 and one establishes a series of relations between the composite NG bosons
 and the fundamental fermionic degrees of freedom. In application to light-cone theory, this 
standard procedure fails due to the non-locality of the light-cone Hamiltonian. The infinite 
range of the non-locality induced by the inverse of $p_-$ destroys the connection between
 soft modulations and low excitation energies. This is akin to the suppression of NG 
 particles in the presence of long-range interactions in gauge theories. We will  demonstrate
  this failure of the standard procedure and show that the excitation of massive particles
 is not suppressed in the limit of long-wavelength modulations. In order to adjust the procedure
 to the  light-cone Hamiltonian non-local transformations of the fermion fields have to be
 considered. We will carry out such an analysis by deriving the general form of the light-cone
 WT identities and by explicitly constructing  the non-local symmetry transformations for 
various model field theories. In addition to the derivation of Gell-Mann--Oakes--Renner (GOR)  like 
relations \cite{GOR68} we consider the connection between  fermion propagator and the structure 
function of the NG bosons \cite{Miransky93}. 


\section{Fermionic Light-Cone Field Theories} 
In this section we will consider fermionic theories described by a Lagrangian of the following structure,
\begin{equation}
  \label{la1}
{\cal L}= \bar{\psi}({\rm i}D_{\mu}\gamma^{\mu}-m)\psi+{\cal L}_{\rm int}
(\bar{\psi},\psi) \, . \end{equation}
The covariant derivative
\begin{equation}
D_{\mu}=\partial_{\mu}+{\rm i}eA_{\mu} 
 \label{cov}
\end{equation}
couples the fermions to a gauge field which might be either external or dynamical. In the latter case 
 integration over the gauge field has to be performed. The Lagrangian (\ref{la1}) may also include a
 fermionic self-interaction, ${\cal L}_{\rm int}$. 
The expression for ${\cal L}$ may contain implicitly sums over  fermion species (``color'') while flavor
dependences important in phenomenological applications are irrelevant for our discussion.
We use a representation of the Dirac matrices particularly suited to the light-cone approach,
\begin{eqnarray}
\alpha_1 &=& \left( \begin{array}{cc} 0 & \sigma_3 \\ \sigma_3  & 0 \end{array}\right)
\, , \quad  \alpha_2 \ =\  \left( \begin{array}{rr} 0 & -{\rm i} \\ {\rm i}  & 0 \end{array}\right)\, ,
\nonumber \\
\alpha_3 &=& \left( \begin{array}{rr} 1 & 0 \\ 0  & -1 \end{array}\right)
\, , \quad  
\beta \ =\  \left( \begin{array}{rr} 0 & - 1 \\ -1   & 0 \end{array}\right) \, .
\label{Diracmat}
\end{eqnarray}
Then, $\gamma_{5}$ and the projection
operators $\Lambda^{\pm}$ are given by
\begin{equation}
\label{gamma}
  \gamma_{5}=\left (\begin{array}{cc} \sigma_{3}&0\\\noalign{\medskip} 0&\sigma_{3}
\end{array}\right ),\  \Lambda^{\pm}=\frac{1\pm \gamma^{0}\gamma^{3}}{2},\
 \gamma^{0}\gamma^{3}=\left (\begin{array}{rr} 1&0\\\noalign{\medskip} 0& -1
\end{array}\right ) .
\end{equation}
The projection operators  $\Lambda^{\pm}$ decompose the 4-spinor into 2-spinors,
\begin{displaymath}
\psi =\frac{1}{2^{1/4}}\left (\begin {array}{c} \varphi\\ \noalign{\medskip}\chi
\end {array}\right )\, ,
\end{displaymath}
and the Lagrangian becomes
\begin{eqnarray}
{\cal L} &=& 
{\rm i}\varphi^{\dagger}D_{+}\varphi+{\rm i}\chi^{\dagger}D_{-}\chi
\nonumber \\
& & +
\frac{{\rm i}}{\sqrt{2}}\left(\chi^{\dagger}
D_{m}\varphi  -  \varphi^{\dagger}D_{m}^{\dagger}\chi \right)+{\cal L}_{\rm int}(\varphi,\chi)
  \label{lc4}
\end{eqnarray}
with
\begin{equation}
  \label{DM}
D_{m}=\sigma_{3}D_{1}+{\rm i}D_{2}-{\rm i}\sigma_{1}m \, .
\end{equation}
Only the spinor $\varphi$ is dynamical since no time derivative of $\chi$ is present. In canonical
quantization, $\chi$ is treated as a constrained field. This reduction in the number of dynamical
degrees of freedom makes the single particle states with given momentum unique and thereby the
light-cone vacuum trivial.
In the representation  (\ref{gamma}), chiral rotations are defined by
\begin{equation}
  \label{chirot}
  \varphi \rightarrow {\rm e}^{{\rm i}\alpha \sigma_{3}}\varphi \, ,\quad \chi \rightarrow
{\rm e}^{{\rm i}\alpha \sigma_{3}}\chi\, .
\end{equation}
The light-cone components of the associated axial current are given by
\begin{eqnarray}
j^{+}_{5}&=&\varphi^{\dagger}\sigma_{3}\varphi \, , \qquad \qquad \ \ \ 
j^{-}_{5}\ =\ \chi^{\dagger}\sigma_{3}\chi \, , \\
j^{1}_{5}&=&\frac{1}{\sqrt{2}}(\varphi^{\dagger}\chi+\chi^{\dagger}\varphi)\, ,\quad
j^{2}_{5}\ =\ \frac{\rm i}{\sqrt{2}}(\chi^{\dagger}\sigma_{3}\varphi-\varphi^{\dagger}\sigma_{3}\chi)\, .
\nonumber
\label{curr}
\end{eqnarray}
We also note that in this representation, the relevant fermion bilinears read
\begin{eqnarray}
  \label{condst}
  \bar{\psi} {\rm i} \gamma_{5}\psi &=&
-\frac{1}{\sqrt{2}}\left(\varphi^{\dagger}\sigma_{2}\chi
  +\chi^{\dagger}\sigma_{2}\varphi\right)\, , \nonumber \\
\qquad \bar{\psi}\psi &=&
-\frac{1}{\sqrt{2}}\left(\varphi^{\dagger}\sigma_{1}\chi
  +\chi^{\dagger}\sigma_{1}\varphi\right)\, .
\end{eqnarray}
With the choice of the four-fermion interaction
\begin{eqnarray}
 \label{ NJL}
 {\cal L}_{\rm int} &=& \frac{g^2}{2}\left[(\bar{\psi}\psi)^{2} +(\bar{\psi}{\rm i}\gamma_{5}
\psi)^2\right] \\
&=& \frac{g^2}{4}\left[\left(\varphi^{\dagger}\sigma_{1}\chi
  +\chi^{\dagger}\sigma_{1}\varphi\right)^2+\left(\varphi^{\dagger}\sigma_{2}\chi
  +\chi^{\dagger}\sigma_{2}\varphi\right)^2\right] \, ,
\nonumber
\end{eqnarray}
the Lagrangian of Eq.~(\ref{la1}) without the gauge field is that of the (one flavour) Nambu--Jona-Lasinio (NJL) 
model \cite{NJL61}. It is
invariant under chiral
rotations provided  the (bare) mass $m$ vanishes.

In  the standard approach to light-cone quantization one employs the canonical formalism after 
eliminating the constrained variables \cite{BPP98}. For investigating issues related to symmetries the path 
integral provides a more appropriate framework. We  start our investigations with the generating functional 
\begin{eqnarray}
  \label{gefu}
  Z [ \eta,\gamma] &=& \int {\rm D} [\varphi, \chi ] \,  \\
& \exp & \!\!\! \left\{ {\rm  i} \int {\rm d}^4x\, \left({\cal L}+
 \eta^{\dagger} \varphi + \varphi^{\dagger} \eta +    \gamma^{\dagger} \chi 
+ \chi ^{\dagger} \gamma  \right)\right\} \, .
\nonumber
\end{eqnarray}
We define fermionic light-cone field theories as effective theories obtained by integrating out the 
``constrained'' fields $\chi$. This is the quantum mechanical version of eliminating the constrained 
variables by solution of the associated Euler-Lagrange equations. In integrating out  $\chi$ all 
quantum fluctuations of these variables are kept. 
Both the kinetic terms and the four-fermion interaction of the NJL model are quadratic in  $\chi$.
Therefore these variables can be eliminated by evaluating  Gaussian integrals with the result
\begin{equation}
  \label{genfu}
  Z[\eta,\gamma] = \int {\rm D}[\varphi]\, \exp \left\{ {\rm i}S[\varphi]+{\rm i}s[\eta,\gamma]  \right\}  \, .
\end{equation}
Action and source terms are given by 
\begin{equation}
  \label{act0}
  S[\,\varphi\,]= \int {\rm d}^4 x \left({\rm i}\varphi^{\dagger}D_{+}\varphi + {\rm i}\chi^{\dagger}
D_{-}[\varphi]\chi\,\right)+{\rm i}\, \mbox{tr}\ln D_-[\varphi] , 
 \end{equation}
\begin{equation}
 \label{source}
s[\eta,\gamma] = \int {\rm d}^{4}x \left(\eta^{\dagger}\varphi+\varphi^{\dagger}\eta +\gamma^{\dagger}\chi
+\chi^{\dagger}\gamma
 +\gamma^{\dagger}\frac{\rm i}{D_{-}[\varphi]}\gamma\, \right) . 
\end{equation}
The action contains the composite field 
\begin{equation}
  \label{chi}
  \chi= -\frac{1}{\sqrt{2}D_{-}[\varphi]}D_{m}\varphi\, . 
\end{equation}
In the generating functional, we have retained  the associated source $\gamma$. 
 The field dependent covariant derivative is given by 
\begin{equation}
  \label{ficov}
D_{-}[\varphi] = D_{-}+{\rm i}g^{2}\left (\begin{array}{rr} \varphi_{2}^{\dagger}\varphi_{2}&\varphi_{2}
\varphi_{1}\\\noalign{\medskip} \varphi_{1}^{\dagger}\varphi_{2}^{\dagger}&- \varphi_{1}^{\dagger}
\varphi_{1}\end{array}\right)  
\end{equation}
and acts on the source $\gamma$ and the spinors 
\begin{equation}
  \label{spin}
\varphi = \left (\begin {array}{c} \varphi_{1}\\ \noalign{\medskip}\varphi_{2}^{\dagger}
\end {array}\right )\quad,\quad \chi  = \left (\begin {array}{c} \chi_{1}\\ \noalign{\medskip}\chi_{2}^{\dagger}\end {array}\right )\, .
\end{equation}
 For most of  our following studies  it will be  convenient  to  eliminate the four-fermion interaction in
 favor of two auxiliary bosonic fields, $\sigma$ and $\pi$. To this end we rewrite the generating functional  (\ref{gefu}) as
\begin{eqnarray}
Z[\eta,\gamma] &=& \int {\rm D}[\varphi,\chi,\sigma,\pi]\, \exp \Biggl\{ {\rm i}S[\varphi,\chi,\sigma,\pi] 
\nonumber  \\
& & 
+{\rm i}\int d^{4}x \left(\eta^{\dagger}
\varphi+\varphi^{\dagger}\eta +\gamma^{\dagger}\chi+\chi^{\dagger}\gamma
  \right)\Biggr\}\, ,
\nonumber
\end{eqnarray} 
with the action 
\begin{eqnarray}
S[\varphi,\chi,\sigma,\pi] &=&
\int {\rm d}^{4}x \Biggl({\rm i}\varphi^{\dagger}D_{+}\varphi+{\rm i}\chi^{\dagger}D_{-}\chi
\nonumber \\
& + &  \frac{\rm i}
{\sqrt{2}}\left(\chi^{\dagger}
D_{\perp}\varphi-\varphi^{\dagger}D_{\perp}^{\dagger}\chi\right) -\frac{1}{2}(\sigma^2+\pi^2) \Biggr)
\nonumber 
\end{eqnarray}
and 
\begin{equation}
  \label{D}
D_{\perp}=\sigma_{3}D_{1}+{\rm i}D_{2}-{\rm i}g(\sigma\sigma_{1}+\pi\sigma_{2})-{\rm i}
m \sigma_{1}\, .
\end{equation}
Integrating out the constrained variables $\chi$ the generating functional becomes 
\begin{equation}
  \label{genfu1}
  Z[\eta,\gamma] = \int {\rm D}[\varphi,\sigma,\pi]\, \exp \left\{ {\rm i}S[\varphi,\sigma,\pi]+{\rm i}s[\eta,\gamma]\right\}\, ,
\end{equation}
with action and  source terms  given by
\begin{equation}
  \label{act}
  S[\varphi,\sigma,\pi]= \int {\rm d}^4 x \left({\rm i}\varphi^{\dagger}D_{+}\varphi+{\rm i}\chi^{\dagger}
D_{-}\chi-\frac{1}{2}(\sigma^2+\pi^2)\right) ,
\end{equation}
\begin{equation}
\label{source1}
s[\eta,\gamma] = \int {\rm d}^{4}x \left(\eta^{\dagger}\varphi+\varphi^{\dagger}\eta +\gamma^{\dagger}\chi+\chi^{\dagger}
\gamma +\gamma^{\dagger}\frac{\rm i}{D_{-}}\gamma \right) .  
\end{equation}
Here the composite field  $\chi$ is defined as
\begin{equation}
  \label{chi1}
  \chi=\chi[\varphi]= -\frac{1}{\sqrt{2}D_{-}}D_{\perp}\varphi\, . 
\end{equation}
In this formulation, the energy of a field configuration $\{\varphi, \sigma, \pi\}$  assumes a particularly
 simple form if expressed in terms of the composite field $\chi$,
\begin{equation}
  \label{HAM}
E = \int {\rm d}^3 x \left({\rm i}\chi^{\dagger}D_{-}\chi+\frac{1}{2}(\sigma^2+\pi^2)\right)\, .
\end{equation}


\section{Local Modulations of Chiral Symmetry Transformations}


\subsection{Local Ward-Takahashi Identities}  
The process of eliminating the constrained degrees of freedom has not altered the symmetry properties
 of the system. In the chiral limit, the action  $S[\varphi]$ [Eq.~(\ref{act0})] is invariant under the global transformation
\begin{equation}
  \label{crot2}
\varphi\rightarrow {\rm e}^{{\rm i}\alpha \sigma_{3}}\varphi  
\end{equation}
since both $D_{-}[\varphi]$ and $D_m$ are invariant and therefore 
$$\chi \rightarrow {\rm e}^{{\rm i}\alpha\sigma_3}\chi \, . $$  
Likewise the invariance of the action $S[\varphi,\sigma,\pi]$  [Eq.~(\ref{act})]
under the combined chiral rotation 
\begin{equation}
  \label{ct0} 
\varphi \rightarrow {\rm e}^{{\rm i}\alpha \sigma_{3}}\varphi \, , 
\end{equation} 
\begin{eqnarray}
\sigma  & \rightarrow & \tilde{\sigma}=\sigma\cos 2\alpha +\pi\sin 2\alpha \, ,
\nonumber \\
 \pi 
&\rightarrow& \tilde{\pi}= \pi\cos 2\alpha -\sigma\sin 2\alpha \, ,
\label{ct}   
\end{eqnarray} 
is easily established.
We also note that the expression (\ref{curr}) for the axial current remains valid, since the
Lagrangian does not contain derivatives of the auxiliary fields. 

The consequences of the presence of a symmetry, in particular the issue of spontaneous symmetry
 breaking, are in general studied by extending the global symmetry transformations characterized by
  $\alpha$ to local ones determined by a space-time dependent symmetry parameter $\alpha(x)$ \cite{Miransky93}. 
In general, such transformations do not leave  the Lagrangian invariant. The  non-invariance is 
connected to the space-time derivatives of the fields in the Lagrangian.  In ordinary coordinates
 with the local form of the Lagrangian the non-invariance can therefore be expressed in terms of 
space-time derivatives of the symmetry parameter  and tends to zero when the  wavelength of $\alpha(x)$
 approaches infinity. It is therefore plausible that in local formulations of field theory the 
space-time modulated generator of the symmetry, {\em e.g.} the operator
\begin{equation}
  \label{op5}
  U_{5}[\alpha] =\exp\left({\rm i}\int {\rm d}^3 x \, \psi^{\dagger}(x)\gamma_{5}\alpha(x)\psi(x)\right)
\end{equation}
in chirally symmetric theories, 
generates in the symmetry-broken case the soft modulations around the vacuum which can be
 interpreted as the associated NG particles. This method of local extension of the symmetry
 transformations is also the basis of the WT  identities \cite{Miransky93}. If for instance we carry out
 in the generating functional (\ref{gefu}) the variable substitution
 $$   \varphi(x)\rightarrow {\rm e}^{{\rm i}\alpha(x) \sigma_{3}}\varphi(x)\, ,\quad \chi(x)\rightarrow
{\rm e}^{{\rm i}\alpha(x) \sigma_{3}}\chi(x)  $$
(replacing covariant derivatives by ordinary derivatives)
the invariance of the path integral yields the functional identity
\begin{eqnarray}
\label{funid}
0 &=& \int D[\varphi,\chi]{\rm e}^{{\rm i}\int {\rm d}^{4}x \left({\cal
      L}+\eta^{\dagger}\varphi+\varphi^{\dagger}\eta
    +\gamma^{\dagger}\chi+\chi^{\dagger}\gamma\right)}
\nonumber \\
& & \Big[\partial_{\mu}j_{5}^{\mu}(x)+\sqrt{2}m
\left(\varphi^{\dagger}\sigma_{2}\chi +\chi^{\dagger}\sigma_{2}\varphi\right)(x) \\
& & +{\rm i}\left(\eta^{\dagger}\sigma_{3}\varphi-\varphi^{\dagger}\sigma_{3}\eta+\gamma^{\dagger}
\sigma_{3}\chi-\chi^{\dagger}\sigma_{3}\gamma\right)(x) \Big]. \nonumber 
\end{eqnarray}
By functional differentiation with respect to the sources the various chiral WT identities can be derived. 
For instance, application of 
\begin{equation}
  \label{sig2}
  \Delta(y)= \sigma_{2}^{\beta\alpha}\left[\frac{\delta}{\delta
  \gamma_{\alpha}^{\dagger} (y)}\frac{\delta}{\delta
  \eta_{\beta}(y)}+\frac{\delta}{\delta
  \eta_{\alpha}^{\dagger} (y)} \frac{\delta}{\delta
  \gamma_{\beta}(y)}\right]_{\eta=\gamma=0}
\end{equation}
yields the WT identity leading to the GOR relation 
\begin{eqnarray}
\label{GMO}
0 &=& \int {\rm D}[\varphi,\chi]\, {\rm e}^{{\rm i}\int {\rm d}^{4}x {\cal
      L}}\Bigg[\left(\varphi^{\dagger}\sigma_{2}\chi
  +\chi^{\dagger}\sigma_{2}\varphi\right) (y)
\nonumber \\
& & 
 \bigg\{\partial_{\mu}j_{5}^{\mu}(x)
+\sqrt{2}
m\left(\varphi^{\dagger}\sigma_{2}\chi +\chi^{\dagger}\sigma_{2}\varphi\right) (x) \bigg\}
\nonumber \\
& & +2{\rm i}\delta(x-y)\left(\varphi^{\dagger}\sigma_{1}\chi
  +\chi^{\dagger}\sigma_{1}\varphi\right)(y) \Bigg] . 
\end{eqnarray}
Although formulated in terms of the light-cone representation of the spinors, the derivation and 
the result are identical to the standard procedure. In light-cone quantization we deal only with the 
spinors $\varphi$ as dynamical variables or, equivalently, as independent integration variables
 in the corresponding generating functional and derive  WT identities by performing the variable substitution 
 \begin{equation}
   \label{locchir}
\varphi(x)\rightarrow {\rm e}^{{\rm i}\alpha(x) \sigma_{3}}\varphi(x)  \end{equation}
in the relevant path integrals such as  (\ref{genfu}). 
 The induced transformation of the composite field $\chi$ [Eq.~(\ref{chi})] is non-local. Repeating the
 above procedure yields the following  WT identity formulated in terms of the dynamical degrees  of freedom $\varphi$
\begin{eqnarray}
\label{GMO3}
0 &=& \int {\rm D}[\varphi]\, {\rm e}^{{\rm i}\int {\rm d}^{4}x \left({\cal
      L_{\varphi}[\gamma,\gamma^{\dagger}]}+\eta^{\dagger}\varphi+\varphi^{\dagger}\eta\right)}
\Big[\partial_{\mu}j_{5,\gamma}^{\mu}(x)
\nonumber \\
& & +\sqrt{2}m\left(\varphi^{\dagger}\sigma_{2}\chi_{\gamma} 
+\chi_{\gamma}^{\dagger}\sigma_{2}\varphi\right)(x)   \\
& & +{\rm i}\left(\eta^{\dagger}\sigma_{3}\varphi-\varphi^{\dagger}\sigma_{3}\eta+\gamma^{\dagger}
\sigma_{3}\chi_{\gamma}-\chi_{\gamma}^{\dagger}\sigma_{3}\gamma\right)(x)  \Big]
\, , \nonumber
\end{eqnarray}
with
\begin{displaymath}
\chi_{\gamma} = \chi+\frac{\rm i}{D_-}\gamma\, ,\quad j_{5,\gamma} =
j_{5}[\varphi,\varphi^{\dagger},\chi_{\gamma},\chi_{\gamma}^{\dagger}]\, .
\end{displaymath}

Integration over the constrained variables has led to a source dependent modification of the
 composite field which accounts for the fluctuations of $\chi$. The WT identity for the $\varphi$
 variables is equivalent to the WT identity formulated in terms of the functional integral over
 $\varphi$ and $\chi$ fields. In the present form, the WT identity has a canonical interpretation
 in light-cone quantization. GOR type relations can be derived from Eq.~(\ref{GMO3}) by functional 
differentiation with respect to the sources.


\subsection{Failure of Pion Dominance}

Chiral WT identities are most useful if one considers their Fourier transform in the limit $p\to 0$.
In normal coordinates this enables us to derive the well-known GOR relation for the pion mass 
away from the chiral limit \cite{GOR68} as well as to relate the pion Bethe-Salpeter (BS) amplitude to the 
quark propagator in the chiral limit \cite{Miransky93}. Being dominated by symmetry aspects such relations have
a universal character independent of the specific dynamics and are therefore of fundamental 
interest. The point $p=0$ is singled out here since only at these kinematics a certain sum over
many meson states appearing in the WT identity is saturated by the NG boson. In view of the
notorious light-cone singularity at $p=0$, it is not obvious whether the same procedure can be 
carried over to light-cone field theory. 
In this section we will show that this is indeed not the case: If one takes the limit
$p_+,p_{\perp}\to 0$ keeping $p_-$ finite, the sum over states is not dominated by the pion
contribution but massive states can come in. We will also recall why this does not happen in ordinary coordinates.
Once we have identified 
the source of the problem we will be able to cure it in later sections by an appropriate
modification of the WT identities.    
  
Our starting point is the WT identity (\ref{GMO3}) which is needed here for $\gamma=0$ only.
We apply the functional derivative
\begin{equation}
\delta_{\alpha \beta}(y,z) \equiv \left.\frac{\delta^2}
{\delta \eta_{\alpha}^{\dagger}(y) \delta \eta_{\beta}(z)}\right
\vert_{\eta=\gamma=0}
\label{d11}
\end{equation}
and obtain 
\begin{eqnarray}
\label{wt12}
0 &=& \int {\rm D} [ \varphi]\, {\rm e}^{{\rm i}S}  
\Biggl\{ \Bigl[ \partial_{\mu}j_{5}^{\mu}(x)+\sqrt{2}m
(\varphi^{\dagger}\sigma_{2}\chi 
\nonumber \\
& + &  
\chi^{\dagger}\sigma_{2}\varphi)(x)\Bigr]
\varphi_{\beta}^{\dagger}(z)\varphi_{\alpha}(y) +
\delta(x-y)\varphi_{\beta}^{\dagger}(z)
(\sigma_3\varphi)_{\alpha}(x)
 \nonumber \\
  & - &   
\delta(x-z)(\varphi^{\dagger}\sigma_3)_{\beta}(x)
\varphi_{\alpha}(y)  \Biggr\} \, .  
\end{eqnarray}
The canonical form of Eq.~(\ref{wt12}) in the chiral limit is 
\begin{eqnarray}
\label{wt13}
& \partial_{\mu}&\!\!\! \langle 0|T(j^{\mu}_{5}(x)\varphi_{\beta}^{\dagger}(z)
\varphi_{\alpha}(y))| 0 \rangle
 \\
& & \qquad =  {\rm i}\left[\delta(x-y)-\delta(x-z)\right](\sigma_3)_{\alpha \beta}G(y-z) \, ,
\nonumber
\end{eqnarray}
where $G$ stands for the light-cone fermion propagator
\begin{equation}
\langle 0|T(\varphi_{\alpha}(y)\varphi_{\beta}^{\dagger}(z))|0 \rangle = 
-{\rm i} \delta_{\alpha \beta}G(y-z)\, .
\label{prop1}
\end{equation}
In the chiral limit the Fourier transform of Eq.~(\ref{wt13}), using 
\begin{eqnarray}
\label{wt14}
F_{5 \alpha \beta}^{\mu}(p,q) & \equiv& \int {\rm d}^{4}x \, {\rm d}^{4}y \,{\rm e}^{-{\rm i}(px-qy)}
\nonumber \\
& & 
\langle 0|
T(j^{\mu}_{5}(x)\varphi_{\beta}^{\dagger}(0)\varphi_{\alpha}(y))|0 \rangle \, ,
\end{eqnarray}
assumes the compact form 
\begin{equation}
\label{wt15}
p^{\mu}F_{5}^{\mu}(p,q)=\sigma_3 \left
(G(q-p)-G(q)\right)\, .
\end{equation}
To calculate the vertex function (\ref{wt14}) by inserting a complete set of states,
we choose the kinematical condition $p_{-}>q_{-}>0 $.
Then, only the time ordering $x^{+}<(y^{+},0)$  contributes and we obtain 
\begin{eqnarray}
  p_{\mu}F^{\mu}_{5 \alpha \beta}(p,q) 
&=& \displaystyle \sum_{n} \frac{p_{\mu}}{2p_{-}}\int {\rm d}x^{+}\int {\rm d}^{4}y 
\nonumber \\
& & 
{\rm e}^{-{\rm i}\left(p_{+}-p_{+}^{(n)}\right)x^{+}+{\rm i}qy}\theta(y^{+}-x^{+})\theta(-x^{+}) 
\nonumber \\
& & \langle 0|T(\varphi_{\beta}^{+}(0)\varphi_{\alpha}(y))
|n;\vec{p}\,\rangle \langle n; \vec{p}\,|j^{\mu}_{5}(0)|0 \rangle\, ,
\nonumber
\end{eqnarray}
where $|n;\vec{p}\, \rangle$ is a mesonic state with mass $m_{n}$ and 
the light-cone momenta are 
\begin{equation}
 \vec{p}=(p_-,\vec{p}_{\perp}) \ , \qquad   p^{(n)}_+=\frac{m_n^2 + p^2_{\perp}}{2p_{-}}\, .
\label{pnplus}
\end{equation} 
The integration over $x^{+}$ finally yields
\begin{eqnarray}
 p^{\mu}F^{5}_{\mu \alpha \beta}(p,q)  
&=& {\rm i}\displaystyle \sum_{n}\frac{p_{\mu}\langle n;\vec{p}\,|j^{\mu}_{5}(0)
|0 \rangle}{p^{2}-m_{n}^{2}+{\rm i}\epsilon} \int {\rm d}^{4}y  \, {\rm e}^{{\rm i}qy}
\nonumber \\
& &  \left\{ \theta(-y^{+})
\langle 0|\varphi_{\beta}^{+}(-y)\varphi_{\alpha}(0)|n;\vec{p}\, \rangle
e^{-{\rm i}py} \right. \nonumber \\  
& & - \left.\theta(y^{+})\langle 0|\varphi_{\alpha}(y)\varphi_{\beta}^{+}(0)
|n;\vec{p}\, \rangle \right\}.
\label{wt17}  
\end{eqnarray}
Up to this point, the calculation in standard  and in light-cone coordinates are essentially 
identical. Differences show up in the attempt to project out the contribution from the NG boson
in the sum over $n$. To identify the origin of these differences, let us consider  in some detail
the expression $p_{\mu}\langle n; \vec{p}\,|j^{\mu}_{5}(0)|0 \rangle$, treating ordinary
and light-cone coordinates in parallel.
Due to covariance, only pseudoscalar (PS) or axial vector (AV) states can contribute to the relevant current 
matrix  element,  
\begin{equation}
\langle n;\vec{p}\, | j_5^{\mu}(0) |0\rangle = \left\{ \begin{array}{ll}
2 f_n p^{(n)}_{\mu} & {\rm PS} \\ 2\left( f_n p^{(n)}_{\mu} + g_n \varepsilon_{\mu}\right) & {\rm AV}
\end{array} \right.
\label{ins1}
\end{equation}
where $\varepsilon$ is a polarization vector and $p^{n} =(\sqrt{m_n^2 + p^2}, \vec{p}\,)$
replaces (\ref{pnplus}) in ordinary coordinates. Current conservation then implies
\begin{equation}
0 = p_{\mu}^{(n)}\langle n;\vec{p}\, | j_5^{\mu}(0) |0\rangle = \left\{ \begin{array}{ll}
2 f_n m_n^2 & {\rm PS} \\ 2\left( f_n m_n^2 + g_n \varepsilon p^{(n)} \right) & {\rm AV}
\end{array} \right.
\label{ins2}
\end{equation}
From this we conclude that $f_n=0$ for all massive PS states which therefore cannot contribute
to the sum over $n$, independently of the coordinates chosen.
For massive AV states, since $\varepsilon p^{(n)}=0$,  
we can only infer that $f_n=0$ whereas $g_n\neq 0$ in general, so that they could in principle 
contribute. 
The further discussion requires specification of the polarization vectors. In ordinary
coordinates, the conditions $\varepsilon^2=-1$ and $\varepsilon p^{(n)}=0$ can be fulfilled by
transverse or longitudinal polarization vectors as follows,
\begin{eqnarray}
\varepsilon_{t} & = & (0, \vec{n}_{\perp} ) \nonumber \\
\varepsilon_{l} & = & \left( \frac{|\vec{p}\,|}{m_n} , \frac{p_0^{(n)}}{m_n} \frac{\vec{p}}{|\vec{p}\,|}\right)
\label{ins3}
\end{eqnarray}
($\vec{n}_{\perp}$ is a unit vector orthogonal to $\vec{p}$\,).
The light-cone components of $\varepsilon$ are then defined in the usual way.
The quantity which enters the WT identity (\ref{wt17}) is $p_{\mu} \langle n;\vec{p}\,| j_5^{\mu}(0)|0\rangle$
which is in general non-vanishing not only for the massless PS but also for massive AV states. In ordinary coordinates,
the unwanted AV states can be suppressed by going to the point $p_{\mu}\to 0$,
\begin{eqnarray}
p_{\mu} \langle n;\vec{p}\, |j_5^{\mu}(0)|0\rangle &=& \left(p_{\mu}-p_{\mu}^{(n)}\right) \langle n;\vec{p}\, |j_5^{\mu}(0)|0\rangle
\nonumber \\
& = &  (p_{0}-p_{0}^{(n)}) 2 g_n \varepsilon^0  \nonumber \\
& \begin{array}[t]{c} \to \\
\raisebox{1ex} {\mbox{${\scriptstyle p_{\mu} \to \, 0}$}} \end{array}
& 0 \qquad {\rm AV}
\label{pv3a}
\end{eqnarray}
We have used current conservation and the fact that $\varepsilon^0(\vec{p}=0)=0$, Eq.~(\ref{ins3}).
In light-cone coordinates, we approach the point $p_+=0,p_{\perp}=0$ keeping $p_- \neq 0$
and find 
\begin{eqnarray}
p_{\mu} \langle n; \vec{p}\, | j_5^{\mu}(0) |0\rangle & = & (p_+-p^{(n)}_+)  \varepsilon^+
\\
& \begin{array}[t]{c} \to \\
\raisebox{1ex} {\mbox{${\scriptstyle (p_+,\, \vec{p}_{\perp}) \to \, 0}$}}
\end{array}
&  - p_+^{(n)} \varepsilon^+ \neq 0 \qquad {\rm AV} \nonumber
\label{pv5a}
\end{eqnarray}
since $\varepsilon^+$ does not vanish at this kinematical point.
Thus   the characteristic light-cone  kinematics prevents dominance of the  NG boson in the GOR type  relation (\ref{wt13}).
More generally, this example demonstrates that indeed soft, local modulations of the form (\ref{locchir}) do not excite
exclusively massless states.  


\section{Non-Local Modulations of Symmetry Transformations}


\subsection{Non-Local Ward-Takahashi Identities}

Only in a local formulation of field theory, space-time modulated symmetry transformations as considered
 in Sec. III can be expected to give rise to small excitation energies  for sufficiently long wavelengths of the 
modulation. In the non-local formulations obtained by integrating out the fields $\chi$, a chiral
 transformation with a space-time dependent parameter $\alpha(x)$ cannot be expected to generate excitations of
 low light-cone energy only. For illustration, we calculate the change in the energy (\ref{HAM}) under the transformation
(\ref{ct0},\ref{ct}) with a space-time dependent $\alpha(x)$. With 
\begin{eqnarray}
  \label{dt}
D_{\perp}[\tilde{\sigma},\tilde{\pi}] &=& {\rm e}^{{\rm i}\alpha\sigma_{3}}D_{\perp}[\sigma,\pi]{\rm e}^{-{\rm i}
\alpha\sigma_{3}} +
({\rm i}\partial_{1}-\sigma_{3}\partial_{2})\alpha
\nonumber \\
& & + {\rm i}m \left[\sigma_{1}(\cos 2\alpha-1)-\sigma_{2}\sin 2\alpha\right]\, ,
\end{eqnarray}   
we find in the chiral limit and for infinitesimal, $x_{\perp}$ independent $\alpha(x)$
\begin{eqnarray}
  \label{dlte}
\delta E &=& {\rm i}\int {\rm d}^3 x \left( \delta\chi^{\dagger}D_{-} \chi+\chi^{\dagger}D_{-} \delta\chi \right)  
\nonumber \\
& =&   \int 
{\rm d}^3 x \,\varphi^{\dagger}D_{\perp}^{\dagger} \left[\frac{1}{D_{-}},\alpha\right]\sigma_{3}D_{\perp}\varphi\, .  
\end{eqnarray}
The change in energy induced by weakly modulated $\alpha(x)$ does not approach smoothly the limit
 $\delta E=0$ for constant $\alpha(x)$.    As a consequence we cannot expect excitations generated 
by the local extension of the symmetry transformations [Eq.~(\ref{locchir})] to excite preferentially the 
NG bosons. Indeed we have seen in Sec. III B that GOR type of relations derived from the WT identity 
(\ref{GMO3}) are not saturated in the limit of vanishing momenta ($p_+,p_{\perp}$) by the NG bosons. 
The connection of large excitation energies with long-wavelength phenomena is similar to the Higgs
 mechanism where due to the long range nature of the Coulomb interaction gapless excitations are
 prevented altogether from appearing. Unlike in the case of the Higgs mechanism and despite the light-cone
 non-localities
 NG bosons must exist if the symmetry is spontaneously broken. Their representation in terms
 of local space-time modulated symmetry transformations of the basic fields $\varphi$ is however lost.  
 The expression for the change in energy  (\ref{dlte}) rather suggests to construct transformations which
 are local space-time modulated transformations of the composite field  $\chi$ and not of the fundamental field $\varphi$.  

Before we proceed with the explicit construction of such transformations in specific theories, we derive, 
independently of the particular dynamics,  the general structure of WT identities following from the existence 
of massless particles.  On the light-cone, formation of NG bosons is connected to the appearance 
of a new symmetry property.    The (Goldstone) one-particle states with vanishing transverse momenta are 
degenerate with the ground state. As a consequence an operator  $\Phi(x^+,x^-) $, the field operator of  the 
NG particles with vanishing transverse momentum,  must exist with the property
\begin{equation}
  \label{creat}
  [\Phi(x^+,x^-),H]|0\rangle = 0\, , \ \mbox{\em i.e.}\ ,\ \frac{\partial}{\partial x^+}\Phi(x^+,x^-)|0\rangle = 0\, .
\end{equation}
This equation can be rewritten formally as a   WT identity 
\begin{eqnarray}
  \label{corr1}
  & \frac{\partial}{\partial x^+}& \!\!\!\langle 0|T\left(\Phi(x^+,x^-){\cal O}(y)\right)|0\rangle =
\nonumber \\
 & & \qquad \delta(x^+-y^+)\langle 0|\left[\Phi(x^+,x^-),
{\cal O}(y)\right]|0\rangle\, ,
\end{eqnarray}
where  ${\cal O}(y) $ is an operator local in light-cone time but otherwise arbitrary.
In terms of a functional integral, this WT identity reads
\begin{eqnarray}
  \label{corr2}
&0 & = \ 
  \frac{\delta}{\delta\omega(y)}\int {\rm D}[\varphi] {\rm e}^{{\rm i}S +{\rm i}s}\Bigg[ \frac{\partial}{\partial x^+}\Phi(x^+,x^-) 
\\
&- & \delta(x^+-y^+)
\left(\Phi(x^+ +\epsilon^+,x^-)-\Phi(x^+ -\epsilon^+,x^-)\right)\Bigg]_{\epsilon=0}\nonumber
\end{eqnarray}
where
$$   \frac{\delta}{\delta\omega(y)}{\rm e}^{{\rm i}s} = {\cal O}(y) {\rm e}^{{\rm i}s}   $$
and the limit $\epsilon^+ \rightarrow 0$ has to be taken after functional differentiation. 
We emphasize that the identities (\ref{corr1}, \ref{corr2}) follow from the mere existence of massless particles. 
Their connection  to symmetry properties is not specified though. It has
to be established within a particular dynamical context.

We will first perform such an analysis for field theories like the NJL  model, where  the spontaneous 
breakdown of chiral symmetry occurs by fermion mass generation. In this case, the differential 
operator $D_{\perp}$ [Eq.~(\ref{D})] must contain the dynamically generated fermion mass $M$. In the presence of the
 the four-fermion interaction this  mass will be given in terms of the expectation 
values of the auxiliary fields  
\begin{equation}
  \label{spbr}
  \langle \sigma \rangle = \frac{M}{g},\quad   \langle \pi \rangle = 0 \, ,
\end{equation}
implying a non-vanishing chiral condensate $\langle 0|\bar{\psi}\psi|0\rangle $.
At first sight, a condensate would seem to be incompatible with the triviality of the light-cone vacuum.
In the context of various field-theoretic models, it has been demonstrated that this is not necessarily the 
case \cite{R18,R19}. If one defines expectation values of bilinear fermion operators more carefully via point-splitting
 in light-cone
time, dynamical information about the specific theory comes in and can be shown to reproduce the
same results as in ordinary coordinates. Equivalently, one should interpret operators which
acquire vacuum expectation values ({\rm e.g.} order parameters) canonically as equal $x^+$ limits of Heisenberg
operators rather than Schr\"odinger operators. Similar considerations apply whenever an operator is replaced 
by a $c$-number, for instance in the relation between $D_{\perp}$ and $D_M$ below. 
  
To study the consequences of this assumed breakdown of the chiral symmetry, we consider the following
 non-local transformation of the fundamental fields $\varphi(x)$
\begin{equation}
\varphi(x) \rightarrow \tilde{\varphi}(x)= {\rm e}^{{\rm i}\alpha(x)\sigma_{3}}D_{M}^{-1}{\rm e}^{-{\rm i}
\alpha(x)\sigma_{3}} D_{M}\varphi(x)
\label{nloc}
\end{equation}
where $D_M$ has been defined in Eq.~(\ref{DM}). 
The  transformation parameter $\alpha(x)$ is chosen to be independent of the transverse coordinates $(x^1, x^2)$.
In this case,  the  transformation becomes unitary
and the Jacobian associated with the variable substitution (\ref{nloc}) is one. 
The transformation (\ref{nloc}) acts non-trivially in the broken phase when the dynamical
fermion mass $M$ is non-vanishing, otherwise it reduces to unity.
In the presence of the four-fermion interaction described by the auxiliary fields $\sigma$ and $\pi$,
the transformation (\ref{nloc}) is always meant to be applied simultaneously with the rotation (\ref{ct}) of the
auxiliary fields.
The rationale behind this unfamiliar construction is the following: The dependent and independent
spinors are related by
\begin{equation}
\chi = - \frac{1}{\sqrt{2} D_-} D_{\perp} \varphi \, .
\label{f1}
\end{equation}
The standard global chiral transformation (at $m=0$) is
\begin{eqnarray}
\varphi \  \to \ \tilde{\varphi} &=& U \varphi \, , \nonumber \\
D_{\perp}[\sigma, \pi]\   \to \  D_{\perp}[\tilde{\sigma},\tilde{\pi}] &=& {\rm e}^{{\rm i}\alpha}
D_{\perp}[\sigma,\pi] {\rm e}^{-{\rm i}\alpha}
\label{f2}
\end{eqnarray}
with 
\begin{equation}
U={\rm e}^{{\rm i}\alpha} 
\label{f3}
\end{equation}
and induces the transformation 
\begin{equation}
\chi \ \to \  \tilde{\chi}\  = \ {\rm e}^{{\rm i}\alpha }\chi 
\label{f4}
\end{equation}
of the dependent field $\chi$. Here we modify this transformation such that for weakly $(x^+,x^-)$-dependent  $\alpha$ the
 energy Eq.~(\ref{HAM}) remains unchanged. This is achieved with the following choice of  $U$, 
\begin{equation}
U= {\rm e}^{{\rm i}\alpha \sigma_3} \frac{1}{D_{\perp}} {\rm e}^{-{\rm i}\alpha \sigma_3}D_{\perp}\, .
\label{f5}
\end{equation} 
 By construction both $\chi$ in Eq.~(\ref{f1}) and $\sigma^2+\pi^2$ are invariant.
The transformation (\ref{nloc}) 
follows if we we  approximatively identify  $D_{\perp}$ with $D_M$.  

Returning to the transformation (\ref{nloc}), 
the definition of the composite field  $\chi$, Eq.~(\ref{chi}), then entails the following transformation 
\begin{eqnarray}
  \label{nlocchi}
\chi  \rightarrow  \tilde{\chi}&=&- \frac{1}{\sqrt{2}D_{-}}D_{\perp}[\tilde{\sigma},\tilde{\pi}]\tilde{\varphi}\nonumber\\& =
 & \frac{1}{D_{-}}\Bigg({\rm e}^{{\rm i}\alpha\sigma_{3}}\Omega {\rm e}^{-{\rm i}
\alpha\sigma_{3}}\Omega^{-1}D_{-}\chi 
\nonumber \\
& & 
- \frac{{\rm i}m}{\sqrt{2}} \left[\sigma_{1}(\cos 2\alpha-1)-\sigma_{2}\sin 2\alpha\right]\tilde{\varphi}\Bigg)  
\end{eqnarray}
where we have defined 
\begin{equation}
\label{om}
\Omega[\sigma,\pi] =D_{\perp}[\sigma,\pi]\,D_{M}^{-1} \, .  
\end{equation}
We now proceed by neglecting the fluctuations of the auxiliary fields $\sigma$ and $\pi$ around their
 vacuum expectation values (\ref{spbr}), {\em i.e.}, we replace these fields by their stationary point in the functional 
integral. In this approximation we can identify the differential operators $D_{\perp}$ [Eq.~(\ref{D})] and $D_{M}$
[Eq.~(\ref{DM})], so that $\Omega[\sigma,\pi]\to 1$.
In the chiral limit, $\chi$ then remains invariant [cf. Eq.~(\ref{nlocchi})] while the
elementary field $\varphi$ transforms  non-trivially
in the broken phase where $\sigma$ develops a finite expectation value $M$.
As pointed out above also the energy of a field configuration [Eq.~(\ref{HAM})] remains invariant under 
the  $x^-$-dependent symmetry transformations
and therefore the transformation defined by Eqs.~(\ref{nloc}) and (\ref{ct}) generates massless excitations
with  arbitrary values of the momentum component $ p_{-}$. 

The same conclusion may also be reached by considering the WT identity associated
with the transformation (\ref{nloc}). To this end, we consider the infinitesimal
version of the transformation (\ref{ct}) and (\ref{nloc}) with $x_{\perp}$-independent $\alpha$. 
We then obtain
\begin{equation}
  \label{dphi}
  \delta \varphi(x) =  {\rm i} \alpha(x^+,x^-)\Sigma_3 \varphi(x)
\end{equation}
with 
\begin{equation}
\Sigma_3 = D_M^{-1} \left[ \sigma_3, D_M \right] = \frac{2 M^2 \sigma_3 - 2 {\rm i}M \vec{\sigma}_{\perp} \vec{\partial}_{\perp}}
{-\Delta_{\perp} + M^2} \ .
\label{dphi1}
\end{equation} 
The operator $\Sigma_3$ coincides with the pion operator at zero transverse momentum which can be 
derived in the NJL model by solving the light-cone BS equation explicitly in ladder approximation, see the
Appendix. 
To leading order in the expansion of 
$\Omega$ [Eq.~(\ref{om})], $\chi$ then transforms only in the presence of a non-vanishing bare quark mass, 
\begin{equation}
  \label{dchi}
  \delta \chi ={\rm i}\alpha(x^+,x^-)\frac{\sqrt{2}m}{D_-}\sigma_{2}\, \varphi \, .
\end{equation}
Performing this variable
substitution in the expression (\ref{act}) for  
the action, we obtain
\begin{eqnarray}
 0&=&  \int {\rm D}[\varphi]e^{{\rm i}S[\varphi,\sigma,\pi]+{\rm i}s[\eta,\gamma]}\int {\rm d}^2 
 x_{\perp}
\nonumber \\
& & \Bigl[(D_{+}^{*}\varphi^{\dagger}(x)+{\rm i}\eta^{\dagger}(x))\,( 
\Sigma_{3}\varphi)(x) 
\nonumber \\
& &   +\sqrt{2}m(\chi^{\dagger}\sigma_{2}\varphi)(x)+{\rm c.c.}\Bigr]
  \label{wi4p}
\end{eqnarray}
Applying the differential operator $\Delta(y)$ of Eq.~(\ref{sig2})
we obtain another GOR  like relation
\begin{eqnarray}
\label{GMO2}
&0& =  \int {\rm D}[\varphi]e^{{\rm i}S[\varphi,\sigma,\pi]} \Bigg[\int {\rm d}^2 
x_{\perp}\left(\varphi^{\dagger}\sigma_{2}\chi
  +\chi^{\dagger}\sigma_{2}\varphi\right)(y)
\nonumber \\
& & 
\left\{\partial_{+}
(\varphi^{\dagger}\Sigma_3\varphi)(x) 
+ \sqrt{2}m(\chi^{\dagger}\sigma_{2}\varphi
+\varphi^{\dagger}\sigma_{2}\chi)(x)\right\}
 \\
& & 
-\delta(x^{+}-y^{+})\delta(x^{-}-y^{-})\left(\varphi^{\dagger}\Sigma 
_{3}\sigma_{2}\chi
  -\chi^{\dagger}\sigma_{2}\Sigma_{3}\varphi\right)(y)\Bigg]\, .\nonumber
\end{eqnarray}
We recognize in this expression for the chiral limit the structure of the general WT identity  
(\ref{corr2}) derived from the existence of NG bosons.  We  
have succeeded in connecting the meson field operator with the generator  
of a unitary $x^{\pm}$-dependent transformation  
\begin{equation}
\label{65a}
\Phi(x^+,x^-)= \int {\rm d}^2 x_{\perp}\,  \varphi^{\dagger}\Sigma_{3}\varphi
\end{equation} 
and, in the chiral limit,  Eq.~(\ref{corr2}) coincides with the WT identity 
(\ref{GMO2}) for  the choice
$${\cal O}= \varphi^{\dagger}\sigma_{2}\chi
  +\chi^{\dagger}\sigma_{2}\varphi .$$
In general Eqs.~(\ref{creat}) and (\ref{corr1}) do not hold as operator
 identities. Interactions between the NG particles  as considered
 in the  corrections to the leading order term [see Eq.~(\ref{om})] prevent
 the  vanishing of the commutator in the whole Hilbert space. 

The operator $\Sigma_3$ plays a prominent role here, both as pion operator and as 
the quantity which replaces $\sigma_3$ when going from the generator of local
to the one of non-local chiral transformations on the light-cone. Actually, one can 
understand the structure and role of $\Sigma_3$ rather simply on a purely classical level.
Consider axial current conservation integrated over $x_{\perp}$,
\begin{equation}
\label{cd1}
\int {\rm d}^2 x_{\perp} \partial_{\mu} j_5^{\mu} = \int {\rm d}^2 x_{\perp} \left[
\partial_+ \left( \varphi^{\dagger} \sigma_3 \varphi \right)
+ \partial_- \left( \chi^{\dagger} \sigma_3 \chi \right)\right] = 0  .
\end{equation}
Using the Euler-Lagrange equations
\begin{equation}
\sqrt{2} \partial_+ \varphi = - D_{\perp}^{\dagger}\chi \ , \qquad \sqrt{2} \partial_-\chi  = - D_{\perp} \varphi \, ,
\label{cd2}
\end{equation}
the $\partial_-$ term in Eq.~(\ref{cd1}) can be transformed with the help of
\begin{eqnarray}
\partial_- \left( \chi^{\dagger} \sigma_3 \chi \right) &=&
\varphi^{\dagger} \left( D_{\perp}^{\dagger} \sigma_3 \frac{1}{D_{\perp}^{\dagger}} \right) \partial_+ \varphi
\nonumber \\
& & 
+(\partial_+ \varphi)^{\dagger} \left(\frac{1}{D_{\perp}} \sigma_3 D_{\perp} \right)  \varphi\, .
\label{cd3}
\end{eqnarray}
In the approximation $D_{\perp} \to D_M$ used above, we have
\begin{equation}
D_{\perp}^{\dagger} \sigma_3 \frac{1}{D_{\perp}^{\dagger}}= \frac{1}{D_{\perp}} \sigma_3 D_{\perp} 
\to \frac{D_{M}^{\dagger} \sigma_3 D_{M}}{D_{M}^{\dagger}D_{M}}
\label{cd4}
\end{equation}
independent of $x^+$ so that the $x_{\perp}$-integrated axial current conservation goes over 
into a local ($x^-$-dependent) conservation law,
\begin{equation}
\partial_+ \int {\rm d}^2 x_{\perp}  \varphi^{\dagger} \Sigma_3 \varphi  = 0 \ .
\label{cd5}
\end{equation}
We thus reproduce Eqs. (\ref{creat}) and (\ref{65a}) in the classical limit.
It is instructive to perform the analogous calculation in 1+1 dimensions as well. In a chiral representation
($\gamma^0 =\sigma_1, \gamma^1=-{\rm i}\sigma_2, \gamma_5=\sigma_3$), vector and axial vector
current conservation assume the form
\begin{eqnarray}
\partial_{\mu} j^{\mu} & =&  \partial_+ \left( \varphi^{\dagger} \varphi \right) + \partial_- \left(\chi^{\dagger}\chi\right) = 0  \, ,
\nonumber \\
\partial_{\mu} j^{\mu}_5 & =&  \partial_+ \left( \varphi^{\dagger} \varphi \right) - \partial_- \left( \chi^{\dagger}\chi \right)= 0 \, .
\label{cd6}
\end{eqnarray}
Adding these two equations we get a local conserved quantity, the pion operator, independently
of any dynamics,
\begin{equation}
\partial_+ \left(\varphi^{\dagger} \varphi\right) = 0 \, .
\label{cd7}
\end{equation}
The relation between the appearance of massless particles and a local symmetry in the light-cone approach
to 1+1 dimensional field theories has been noted before \cite{LTLY91}. In the case of the Schwinger model 
\cite{Schwinger62}, the 
axial anomaly generates a mass term, see Sec. IV B.

To analyze further the chiral WT identity (\ref{GMO2}) we proceed as usual by   
 Fourier-transforming the  two point functions  
\begin{eqnarray}
&G_{\Sigma}^+(p) & =\  \int {\rm d}^4 x\, {\rm e}^{{\rm i}px}
\nonumber \\
& & 
\langle 0| 
T(\varphi^{\dagger}(x)\Sigma^{ 
3}\varphi(x)\left(\varphi^{\dagger}(0)\sigma_{2}\chi(0) 
+ {\rm c.c.}\right))|0\rangle \, , 
\nonumber
\end{eqnarray}
\begin{eqnarray}
& G_{5}(p) &=\ \int {\rm d}^4 x \, {\rm e}^{{\rm i}px}
\nonumber \\
& & \!\!\!\!\!\!\!\!\!\!
\langle 0| T((\chi^{\dagger}(x)\sigma_{2}
\varphi(x)+ {\rm c.c.})\left(\varphi^{\dagger}(0)\sigma_{2}\chi(0)+ {\rm c.c.}\right))
|0\rangle 
\nonumber
\end{eqnarray}
and defining the condensate [cf. Eq.~(\ref{condst})]
$$\Gamma = \frac{\rm i}{2\sqrt {2}}\langle 0|\varphi^{\dagger}(y)\Sigma_{3} 
\sigma_{2}\chi(y)
  -\chi^{\dagger}(y)\sigma_{2}\Sigma_{3}\varphi(y)|0\rangle .$$
The WT identity becomes
\begin{equation}
  \label{wid}
 p_{+} G_{\Sigma}^+(p_+,p_-,\vec{0}\,) 
- {\rm i}\sqrt{2}m G_{5}(p_+,p_-,\vec{0}\,) = -{\rm i} \sqrt{2} \Gamma.
\end{equation}
Expressing $G_{\Sigma}^+$ and $G_{5}$ as sums of contributions from mesonic
states as we did in Sec. III B, we have
\begin{equation}
\sum_{n} \frac{p_{+}B_{n}(p_-)C_{n}(p_-) + {\rm i}\sqrt{2}m |C_{n}(p_-)|^2}
{p_{+} - \frac{m_{n}^{2}}{2p_{-}}+{\rm i}\, {\rm sgn}(p_-) \epsilon } = -{\rm i}\sqrt{2} \Gamma  ,
\end{equation}
where $B_{n}$ and $C_{n}$ are defined by
\begin{eqnarray}
B_{n}(p_-) &\equiv &  
\langle n; p_-|\varphi^{\dagger} \Sigma_{3} \varphi|0 \rangle\, , 
\nonumber \\
C_{n}(p_-) &\equiv & 
\langle 0|(\chi^{\dagger}\sigma_{2}\varphi + {\rm c.c.})| n; p_-\rangle \, .
\end{eqnarray}
Examining the above expression near $p_+ = 0$ in the chiral limit, 
we see that a massless meson (pion, $m_{\pi}=0$)
must exist for non-vanishing condensate ($\Gamma$) and obtain
\begin{equation}
B_{\pi}(p_-) C_{\pi}(p_-) = -{\rm i}\sqrt{2}\Gamma \, .
\end{equation}
$B_{\pi}$ is related to the pion decay constant $f_{\pi}$ defined in
Eq.~(\ref{ins1}) as $B_{\pi}(p_-) = 4p_-f_{\pi}$.
For small $m$, near the pion pole, we have
\begin{equation}
m_{\pi}^{2}B_{\pi}(p_-) = {\rm i}2\sqrt{2}p_{-}C_{\pi}^*(p_-)
\end{equation}
which is nothing but the GOR relation
\begin{equation}
m_{\pi}^{2} f_{\pi}^{2} = m \Gamma = -m\langle \bar{\psi}\psi \rangle \, .
\end{equation}  
We have thus shown for the NJL model that 
the modified non-local chiral transformation which leaves the
composite field $\chi$ invariant in the chiral limit generates useful WT identities.
Unlike in Sec. III B, here pion dominance holds on the light-cone,
leading correctly to the massless NG boson and the GOR relation.


\subsection{Quark Propagator and Pion Structure Function}
In ordinary coordinates WT identities have been used to derive a general relation between the BS amplitude
of the pion at total momentum zero on the one hand and the quark propagator on the other hand \cite{Miransky93}.
It is interesting to repeat this derivation in the light-cone approach where the BS amplitude is 
closely related to the structure function of the pion. In the context of strong interaction physics, this opens the possibility
 to relate  quark propagator and  pion  structure function. In ordinary coordinates where the BS amplitude is closely
 related to the form factor of the pion
rather than to the structure function such a connection is much more remote.

Equipped with the tool of ``non-local WT identities" developed in Sec. IV A, we now use this technique to
derive the desired relationship.  
Applying $\delta_{\alpha\beta} (y,z)$ defined in Eq.~(\ref{d11}) to the identity (\ref{wi4p}) in the
chiral limit yields
\begin{eqnarray}
0 & = & \int {\rm D}[\varphi]{\rm e}^{{\rm i}S[\varphi,\sigma,\pi]}
\\
& & 
\biggl\{\int {\rm d}^2x_\perp\partial_+(\varphi^\dagger(x)\Sigma_3\varphi(x))
\varphi_\alpha(y)\varphi^\dagger_\beta(z) \nonumber \\
&&+\delta(x^+-y^+)\delta(x^--y^-)
(\Sigma_3\varphi)_\alpha(y)\varphi_\beta^\dagger(z)
\nonumber \\
& & 
-\delta(x^+-z^+)\delta(x^--z^-)
\varphi_\alpha(y)(\Sigma_3^\ast\varphi^\dagger)_\beta(z)\biggr\} 
\nonumber
\end{eqnarray}
or equivalently the WT identity
\begin{eqnarray}
&\partial_+&\!\!\! \int {\rm d}^2x_\perp
\langle0\vert T(\varphi^\dagger(x)\Sigma_3\varphi(x))
\varphi_\alpha(y)\varphi_\beta^\dagger(z)\vert0\rangle =
\\
&-&\delta(x^+-y^+)\delta(x^--y^-)\langle0\vert
T(\Sigma_3\varphi)_\alpha(y)\varphi_\beta^\dagger(z)\vert0\rangle
\nonumber\\
&+&\delta(x^+-z^+)\delta(x^--z^-)\langle0 
\vert
T\varphi_\alpha(y)(\Sigma_3^\ast\varphi^\dagger)_\beta(z)\vert0\rangle \, .
\nonumber
\end{eqnarray}
With the definition of the three-point function
\begin{equation}
F_{\alpha\beta}(y-x,z-x)
=\langle0\vert T(\varphi^\dagger(x)\Sigma_3\varphi(x))
\varphi_\alpha(y)\varphi_\beta^\dagger(z)\vert0\rangle\, ,
\end{equation}
the WT identity at $x^+=x^-=0$ reads
\begin{eqnarray}
&&\left. \partial_+\int {\rm d}^2x_\perp F_{\alpha\beta}(y-x,z-x) \right|_{x^+=0}
=
\nonumber \\
&& -\delta(y^+)\delta(y^-)
\int {\rm d}^4\xi(y\vert(\Sigma_3)_{\alpha\beta}\vert \xi)G(\xi- 
z)
\nonumber\\
&&
+\delta(z^+)\delta(z^-)\int {\rm d}^4\xi
G(y-\xi)(z\vert(\Sigma_3^\ast)_{\beta\alpha}\vert \xi)  \, ,
\end{eqnarray}
with the fermion propagator  defined in Eq.~(\ref{prop1}).
In terms of the Fourier transformed  three-point function 
\begin{equation}
F(p,q)
=\int {\rm d}^4y{\rm d}^4z\,{\rm e}^{{\rm i}qy+{\rm i}(p-q) z}F(y 
,z)\, ,
\end{equation}
the WT identity in momentum space becomes 
\begin{equation}
p_+F(p,q)
=\Sigma_3(q)\theta(q_-)G(q)-\Sigma_3(q -p)\theta(q_- -p_-)G(q-p) \, .
\end{equation}
The general form of the fermion propagator (in manifestly covariant theories) is
\begin{equation}
G(q)=\frac{2q_-A}{q^2A^2-B^2}=\frac{A}{(q_+-q_\perp^2/2q_-)A^2-B^2/2q 
_-}\, ,
\end{equation}
where $A$ and $B$ are still arbitrary functions of $q^2=2q_+q_--q_\perp^2$.
$\Sigma_3(q)$ is defined by the Fourier transform
\begin{equation}
\Sigma_3(q)=\int {\rm d}^4x\, {\rm e}^{-{\rm i}q x}(0\vert\Sigma_3\vert x)
=\frac{2M}{\sigma_\perp q_\perp+M\sigma_3}\, ,
\label{app1}
\end{equation}
such that for $p_\perp=0$, $\Sigma_3(q-p)=\Sigma_3(q)$.
The one-pion pole term of $F(p,q)$ is given by
\begin{equation}
[F(p,q)]_{\rm pole}
=2f_\pi\frac{1}{p_+}\Psi(p,q)\, ,
\label{pole}
\end{equation}
where we have used the relation between $B_{\pi}(p_-)$ and $f_{\pi}$. 
$\Psi(p,q)$ in (\ref{pole}) is the Fourier transform of
the BS amplitude
\begin{equation}
\Psi_{\alpha\beta}(x,y)
=\sqrt{2p_-}\langle0\vert T\varphi_\alpha(x)\varphi_\beta^\dagger(y)\vert \pi; p
\rangle \, .
\end{equation}
In the one-pion pole approximation and at $p_\perp=0$,  
the BS amplitude is completely determined by the fermion propagator
\begin{equation}
\Psi(p,q)=\frac{1}{2f_\pi}\Sigma_3(q)
\left[\theta(q_-)G(q)-\theta(q_- -p_-)G(q-p)\right]\ .
\label{L77}
\end{equation}
The light-cone wave function for the pion is given by the BS amplitude as
\begin{equation}
\Pi(p_-,q_-,q_{\perp}) = \int {\rm d}q_+ \Psi(p,q) \, ,
\end{equation}
and in turn is related to the pion structure function as
\begin{equation}
{\cal F}\left({\displaystyle\frac{q_-}{p_-}}\right)
=\int {\rm d}^2q_\perp\vert\Pi(p_-,q_-,q_\perp)\vert^2 \, .
\end{equation}
Let us briefly indicate the modifications necessary for the discussion of 1+1 dimensional
field theories below.
For the Gross-Neveu model \cite{GN74} written in terms of auxiliary fields $\sigma$ and 
$\pi$ and with the constrained variable $\chi$  integrated 
out, the  action  is of the same form as that of the NJL model if the composite field 
$\chi$ is defined as
\begin{equation}
\chi=\chi[\varphi]=\frac{1}{\sqrt{2}{\rm i}\partial_-}(m+g\Sigma^\dagger)\varphi\, ,
\qquad\Sigma=\sigma+{\rm i}\pi\ .
\end{equation}
For modulations of the fields in the chiral limit
\begin{equation}
\delta\varphi={\rm i}\alpha(x^+,x^-)\varphi, \  \ 
\delta\Sigma={\rm i}\alpha(x^+,x^-)\Sigma, \ \ 
\delta\chi=0
\end{equation}
we are provided with
\begin{equation}
0=\int {\rm D}[\varphi]{\rm e}^{{\rm i}S[\varphi,\sigma,\pi]+{\rm i}s[\eta,\gamma]}
\{(\partial_+\varphi^\dagger(x)+{\rm i}\eta^\dagger(x))\varphi(x)+{\rm c.c.}\} .
\end{equation}
For the 't~Hooft model \cite{tHooft74} on the other hand, the Lagrangian in the $A_-=0$ gauge is
given by
\begin{eqnarray}
{\cal L} & =& \varphi_i^\dagger({\rm i}\partial_+\delta_{ij}+eA_{+ij})\varphi_j
+{\rm i}\chi_i^\dagger\partial_-\chi_i
\nonumber \\
& & +\partial_-A_{+ij}\partial_-A_{+ji}
-\frac{m}{\sqrt{2}}(\varphi_i^\dagger\chi_i+\chi_i^\dagger\varphi_i).
\end{eqnarray}
$A_{+ij}$ is the solution of Poisson's equation,
\begin{equation}
A_{+ij}=\frac{e}{2}\frac{1}{\partial_-^2}\varphi_j^\dagger\varphi_i \, .
\end{equation}
In the chiral limit, the $\chi_i$ are zero and 
for modulations of the fields 
\begin{equation}
\delta\varphi_i={\rm i}\alpha(x^+,x^-)\varphi_i \, ,
\end{equation}
we obtain once again the same WT identity. The relation between the BS amplitude and 
the fermion propagator can be taken over from the (3+1)-dimensional 
case by simply setting $\Sigma_3 =1$.

We now apply these formulae to concrete models, starting 
from Eq.~(\ref{L77}). 
Consider the NJL model first. To leading order
in the $1/N$ expansion, the Fermion propagator   is just a free, massive Green's function (dynamical mass $M$),
\begin{equation}
G(q) = \left( q_+ -\frac{q_{\perp}^2+M^2}{2q_-} + {\rm i}\,{\rm sgn}(q_-) \epsilon \right)^{-1}\, .
\label{x4}
\end{equation}
Integration over the relative energy variable, 
\begin{equation}
\int {\rm d} q_+ G(q ) = - {\rm i} \pi {\rm sgn}\left( q_- \right) \, ,
\label{x5}
\end{equation}
yields at once the following BS amplitude for the pion,
\begin{equation}
\int {\rm d}q_+ \Psi (p_+,p_-,q) = -\frac{{\rm i}\pi}{2f_{\pi}} \Sigma_3(q_{\perp})
\left[ \theta\left(q_-\right)-\theta\left(q_--p_-\right)\right] \, ,
\label{x6}
\end{equation}
in agreement with the result of the explicit solution of the model on the light-cone (see the Appendix).
Incidentally, had we used the ``local WT identity" of Secs. III A and III B, the result for the 
pion BS amplitude would differ from Eq.~(\ref{x6}) by the replacement $\Sigma_3 \to \sigma_3$ and hence 
give the wrong answer. As explained in Sec. III B this discrepancy is due to the fact that axial vector states
spoil pion dominance in the light-cone calculation. 
In the case of the NJL model, one can actually verify explicitly that the sum over axial vector states
accounts for the difference between $\Sigma_3$ and $\sigma_3$ by making use of  the 
$q\bar{q}$ scattering state solutions of the light-cone Tamm-Dancoff equation. 

For the chiral Gross-Neveu model, the calculation is essentially the same [drop $q_{\perp}^2$
in the propagator, Eq.~(\ref{x4})]. Since $\Sigma_3=1$ in this case, the light-cone wavefunction
reduces to the difference of step functions and becomes constant in the interval $x=p_-/P_-\in
[0,1]$.

In the case of the 't~Hooft model, the quark propagator does not have the standard covariant form.
One finds \cite{tHooft74}
 \begin{equation}
G(q) = \left( q_+ +\frac{N g^2}{2\pi}\frac{1}{2q_-} - \frac{Ng^2}{2\pi \lambda}{\rm sgn}(q_-)
+ {\rm i}\,{\rm sgn}(q_-) \epsilon\right)^{-1}
\label{x7}
\end{equation}
where the limit $\lambda \to 0$ moves the pole to infinity. If we perform the $q_+$ integration
before taking the limit $\lambda \to 0$, the result is the same as in the chiral Gross-Neveu model.
Note that in our derivation, we have assumed a covariant quark propagator, so that the 
final formula strictly speaking cannot be applied to gauge theories. Nevertheless, the
result seems to be more general than the derivation.
The pion wavefunction on the light-cone does not reflect the drastic difference in the fermion
propagators of a confining and a free massive theory, so that quite some information 
seems to get lost when going from the propagator to the structure function due to
the integration over $q_+$.
The simple results for 't~Hooft and Gross-Neveu model
pion wave functions agree with the literature \cite{tHooft74,TO93}.

As a last application of our formalism we consider the Schwinger model \cite{Schwinger62}, paying
special attention to the axial anomaly on the light-cone.
The Lagrangian density in the $A_- = 0$ gauge is
\begin{equation}
{\cal L} = \varphi^{\dagger}\left({\rm i}\partial_+ + eA_+\right)\varphi
+ \frac{1}{2}\left(\partial_-A_+\right)^2.
\end{equation}
To obtain the WT identities with the modulation of the fields
\begin{equation}
\delta \varphi (x) = {\rm i} \alpha (x) \varphi (x),
\end{equation}
we have to consider the variations in the measures of the functional integral 
\begin{eqnarray}
\delta D[\varphi] &=& {\rm i}\int {\rm d}^2 x \, \alpha(x) J(x) D[\varphi]\, , \nonumber \\
\delta D[\varphi^{\dagger}] &=& -{\rm i} \int {\rm d}^2 x \, \alpha(x) \tilde{J}(x)
D[\varphi^{\dagger}]\, .
\end{eqnarray}
$J$ and $\tilde{J}$ are divergent quantities and we regularize them with
the square of the Dirac operator, in analogy with Fujikawa's method \cite{Fujikawa79} in 
ordinary (Euclidean) coordinates. The Dirac operator in the present model
before eliminating the non-dynamical (vanishing) component of the fermion
is a $2\times2$ matrix
\begin{equation}
D \!\!\!\!/ = \left (\begin{array}{cc} 0&\partial_-\\\noalign{\medskip} 
\partial_+ - {\rm i}eA_{+}&0 \end{array}\right )\ \, ,
\end{equation}
and thus $D\!\!\!\!/^{\,2}$ is a diagonal matrix with the upper and lower elements,
$D \!\!\!\!/^{\,2}_+$ and $D \!\!\!\!/^{\,2}_-$ respectively, given by
\begin{equation}
D \!\!\!\!/^{\,2}_+ = \partial_-(\partial_+ - {\rm i}eA_+)\, , \qquad
D \!\!\!\!/^{\,2}_- = (\partial_+ - {\rm i}eA_+)\partial_- \, .
\end{equation}
Since $\varphi$ is the upper component, $J$ is regularized by the
upper diagonal element as
\begin{equation}
J(x) = (x\vert \exp\left\{{\rm i}\frac{D \!\!\!\!/^{\,2}_+}{\Lambda^2}\right\}\vert x) 
     = \frac{\Lambda^2}{2\pi}\left(1 + \frac{e}{2\Lambda^2} \partial_- A_+ (x)\right)\, .
\end{equation}
In accord with the covariant measure ${\rm D}[\psi] {\rm D}[\bar{\psi}]$ used in
ordinary coordinates, $\varphi^{\dagger}$ should be
 considered as the lower component in the present representation and $\tilde{J}$ is to be regularized by the hermitian
conjugate of the lower diagonal element which is equal to the upper element. Thus we
obtain
\begin{equation}
\tilde{J}(x) = (x\vert \exp \left\{-{\rm i}\frac{D \!\!\!\!/^{\,2}_+}{\Lambda^2}\right\}\vert x) 
      = \frac{\Lambda^2}{2\pi}\left(1 - \frac{e}{2\Lambda^2} \partial_- A_+ (x)\right)\, .
\end{equation}
The WT identity obtained by the above modulation can be expressed as
\begin{equation}
\partial_+ \varphi^{\dagger}(x) \varphi(x) = -J(x) + \tilde{J}(x)
= - \frac{e}{2\pi} \partial_- A_+ (x)\, ,
\end{equation}
which represents the anomaly in the Schwinger model. Eliminating the gauge
field, $A_+$, by resolving the Poisson equation
\begin{equation}
\partial_{-}^2 A_+ = e \varphi^{\dagger} \varphi \, ,
\end{equation}
we finally obtain
\begin{equation}
2\partial_- \partial_+ \varphi^{\dagger}\varphi = -\frac{e^2}{\pi}
\varphi^{\dagger}\varphi \, ,
\end{equation}
the Klein-Gordon equation of the Schwinger particle with mass $e/\sqrt{\pi}$.


\section{Conclusions}
We have studied  chiral symmetry in the framework of light-cone field theory. We have analyzed the properties
 of the Nambu-Goldstone realization of the chiral symmetry by formulating appropriate WT identities.
 The non-local form of the light-cone Hamiltonian makes a revision of the standard analysis necessary. We have 
investigated the general structure of the chiral WT identities  and have carried out explicit constructions for the
 NJL model, the Gross-Neveu model, two dimensional QCD and the Schwinger model. Basic
 elements in our construction are independent of the specific dynamics. In particular we have demonstrated
 that derivation of the relevant identities requires consideration of  symmetry transformations which in general 
cannot be generated by soft space-time dependent modulations of the underlying global chiral rotations. Our
 results are a first step towards formulation of the Goldstone theorem in light-cone field theory. As we have 
shown, in light cone field theory, the existence of massless particles leads to a WT identity without reference to
 a spontaneous symmetry breakdown. The connection however of the creation operator of the NG
 particle with symmetry transformations could be established only within a specific dynamical context. In particular
 in the case of the NJL model our detailed construction assumes that fermion mass generation is the basic
 mechanism for the spontaneous chiral symmetry breakdown. Application of  WT identities leads, like in ordinary
 coordinates, to a series of relations connecting properties of the NG bosons with properties of the 
fermionic degrees of freedom. The GOR relation  has  the same structure and physical
 content as the corresponding relation formulated in ordinary coordinates. A novel relation between the 
fermion-propagator and the structure function of the NG bosons has been deduced from the
 light-cone WT identity.\vskip .2cm
An important next step in the study of chiral symmetry breakdown will be the application of the techniques
 developed here to gauge theories. Investigation of the  chiral symmetry breakdown in QED by an external 
magnetic field should be a comparatively straightforward extension of the present investigation. Similarly, 
the extension of our light-cone calculation of the axial anomaly in 1+1 dimensions to 3+1 dimensions should 
be examined. A study of chiral WT identities in light-cone QCD may provide a new perspective on the mechanism
 driving the spontaneous chiral symmetry breakdown. Judging from our treatment of the 't~Hooft model,  fermion 
 mass generation may not be the relevant process. Irrespective of their detailed structure,  these WT identities
 will among others lead to a relation between the pion structure function and the quark propagator, which could
 be of considerable phenomenological interest.

\vskip 1.0cm

\begin{center} {\bf Acknowledgement} \end{center}

K. Y. is grateful for support by an Alexander von Humboldt Foundation Research Award. K. O. and K. Y.
are supported in part by a Grant-in-Aid from the Ministry of Education, Culture, Sports, Science and
Technology of Japan.


\section*{Appendix: NJL Model on the Light-Cone in Tamm-Dancoff Approximation} 
We summarize here the canonical formulation of the (one flavor) NJL model on the light-cone
(see also \cite{BHMY99,Heinzl00,IM01}). 
The action with the auxiliary fields $\sigma$ and $\pi$ integrated out 
can be expressed as
\begin{equation}
S[\varphi] = \int {\rm d}^{4} x \left({\rm i}\varphi^{\dagger} D_{+} \varphi 
+ {\rm i}\chi^{\dagger}\partial_- \chi 
- \frac{1}{2}(\sigma^{2} + \pi^{2})\right) \, ,
\label{1}
\end{equation}
where ${\chi}$ is given in Eq.\ (\ref{chi1}), 
and $\sigma$ and $\pi$ are now the
solutions of the following constraint equations,
\begin{eqnarray}
\sigma + \frac{g}{2}  \varphi^{\dagger} \left( \sigma_1
\frac{1}{\partial_{-}} D_{\perp}
-D_{\perp}^\dagger
\frac{1}{\partial_{-}} \sigma_1 \right) \varphi &=& 0 \, ,
\label{2p} \\
\pi + \frac{g}{2} \varphi^{\dagger} \left(  \sigma_2
\frac{1}{\partial_{-}} D_{\perp}
-D_{\perp}^\dagger
\frac{1}{\partial_{-}} \sigma_2  \right) \varphi &=& 0 \, .
\label{2}
\end{eqnarray}
Eqs.~(\ref{2p}) and (\ref{2}) can be solved for $\sigma$ and $\pi$ 
in a formal $1/N$ expansion by writing
\begin{equation}
\sigma(x)  \approx  \sigma^{(0)} + \sigma^{(1)}(x)\, ,\qquad
\pi(x)  \approx  \sigma^{(2)}(x)\, ,
\label{4}
\end{equation}
where $\sigma^{(0)}$ is the dominant $c$-number part while $\sigma^{(1,2)}$   
describe the fluctuations in the $\sigma$ ($0^+, n = 1$) and $\pi$ ($0^-, n = 2$) channels.
To leading order in this expansion, one can replace $D_{\perp}$ by 
$D_{M}$, cf. Eq.~(\ref{DM}).
The $c$-number part $\sigma^{(0)}$ is determined self-consistently 
via the gap equation
\begin{equation}
\sigma^{(0)} = \frac{1}{2} g \langle 0 |\varphi^\dagger
\sigma_1 \chi+\chi^\dagger\sigma_1\varphi|0\rangle \ ,
\label{5}
\end{equation}
whereas the fluctuations satisfy 
\begin{eqnarray}
& & \sigma^{(n)}(\vec{x}\,) + \int {\rm d}^3y F(\vec{x}-\vec{y}\,) \sigma^{(n)}(\vec{y}\,)
= 
\nonumber \\
& & \!\!\!\!\!\!\!\!\!\! - \frac{g}{2}\int {\rm d}^3y \varphi^{\dagger}(\vec{x}\,)\sigma_n
(\vec{x}\,|\frac{1}{\partial_-}|\vec{y}\,)
D_{M} \varphi(\vec{y}\,) + {\rm h.c.}
\label{11}
\end{eqnarray}
Here, the kernel $F$ is defined as 
\begin{eqnarray}
F(\vec{x}-\vec{y}\,) & \equiv &  - \frac{\rm i}{2} g^2 \left(
\langle \varphi^{\dagger}(\vec{x}\,) \varphi(\vec{y}\,)\rangle
(\vec{x}\,|\frac{1}{\partial_-}|\vec{y}\,) \right.
\nonumber \\
& &  + \left.
\langle \varphi^{\dagger}(\vec{y}\,) \varphi(\vec{x}\,)\rangle
(\vec{y}\,|\frac{1}{\partial_-}|\vec{x}\,) \right) \, .
\label{8}
\end{eqnarray}
Using these equations together with the normal mode expansion of $\varphi$,
\begin{eqnarray}
\varphi_{\alpha}(\vec{x}\,) = \int \frac{{\rm d}^{3}q}{(2\pi)^{3}} 
{\rm e}^{-{\rm i}qx} b_{\alpha}(\vec{q}\,),
\label{3}
\end{eqnarray}
($b_{\alpha}(\vec{q}\,)$ annihilates (creates) a fermion (anti-fermion) 
for positive (negative) $q_{-}$), the Hamiltonian 
up to 2nd order in $1/\sqrt{N}$  
becomes 
\begin{eqnarray}
H  & = & 
\int \frac{{\rm d}^3q}{(2\pi)^3} \frac{q_{\perp}^2+M^2}{2 q_-} b_{\alpha}^{\dagger}
(\vec{q}\,) b_{\alpha}(\vec{q}\,) 
\\
& &  - \frac{1}{2} \int \frac{{\rm d}^3p}{(2\pi)^3} (1+F(p_-))
\sum_{n=1,2} \sigma^{(n)}(\vec{p}\,) \sigma^{(n)}(-\vec{p}\,) \, , \nonumber
\label{15}
\end{eqnarray}
{\em i.e.}, a sum of a free, massive Hamiltonian and a 2-term separable potential.
The formfactors of the interaction term are
\begin{eqnarray}
\sigma^{(n)}(\vec{p}\,) 
& = & \frac{1}{1+F(p_-)}\left(-\frac{g}{2}\right)
\\ & & 
\int \frac{{\rm d}^3 q}{(2\pi)^3}{\cal V}_{\alpha\beta}^{(n)}(\vec{p},\vec{q}\,)
b_{\alpha}^{\dagger}(\vec{q}\,) b_{\beta}(\vec{q}-\vec{p}\,)\, ,  \nonumber 
\end{eqnarray}
where
\begin{eqnarray}
{\cal V}^{(n)}(\vec{p},\vec{q}\,) & = &
\left[ \sigma_n (- (q \!\!\!/_{\perp}- p \!\!\!/_{\perp}) + \sigma_1 M) \right]
\frac{1}{q_--p_-} 
\nonumber \\
& & +
\left[  (- q \!\!\!/_{\perp}^{*}+ \sigma_1 M)\sigma_n  \right]
\frac{1}{q_-}
\label{16}
\end{eqnarray}
with $q \!\!\!/_{\perp} = \sigma_3 q_1 + {\rm i} q_2$ etc. 
In 
Tamm-Dancoff approximation, the meson state vector is expanded as
\begin{equation}
|\sigma^{(n)}; \vec{p}\, \rangle = \int \frac{{\rm d}^3q}{(2\pi)^3}
S_{\alpha \beta}^{(n)} (\vec{p},\vec{q}\,) b_{\alpha}^{\dagger}(\vec{q}\,)
b_{\beta}(\vec{q}-\vec{p}\,) |0 \rangle \, .
\label{18}
\end{equation}
The Schr\"odinger equation within the particle--anti-particle subspace, 
\begin{equation}
\langle 0| b_{\beta}^{\dagger}(\vec{q}-\vec{p}\,) b_{\alpha}(\vec{q}\,) 
(E-H) |\sigma^{(n)}; \vec{p}\, \rangle = 0\, ,
\label{19}
\end{equation}
then gives the Tamm-Dancoff equation
\begin{eqnarray}
 \Bigl(E &-&   \frac{M^2+q_{\perp}^2}{2q_-} -
\frac{M^2+q_{\perp}^2}{2(p_--q_-)}
\Bigr) S^{(n)}(\vec{p},\vec{q}\,)    \nonumber\\
& = & \frac{1}{1+F(q_-)}\left(- \frac{g^2}{2} \right)
{\cal V}^{(n)}(\vec{p},\vec{q}\,) 
 \\
& & 
\int \frac{{\rm d}^3 q'}{(2\pi)^3} {\rm tr}\,
{\cal V}^{(n)}(-\vec{p},\vec{q}\,'-\vec{p}\,)
S^{(n)}(\vec{p},\vec{q}\,')\, . \nonumber
\label{20}
\end{eqnarray}
Since the interaction term on the r.h.s. is separable, the eigenvalue problem
reduces to an algebraic equation which 
for $p_\perp=0$ reads
\begin{eqnarray}
& & 0 = 1 + \frac{1}{1+ \widetilde{F}(p_-)} \frac{g^2}{2} \int \frac{{\rm d}^3q}
{(2\pi)^3} \frac{\theta(q_-) \theta(p_--q_-)}
{E- \frac{M^2+q_{\perp}^2}{2q_-} - \frac{M^2+q_{\perp}^2}{2(p_--q_-)}}
\nonumber \\
& &
\left[ (q_{\perp}^2 + M^2) \left( \frac{1}{q_-}+ \frac{1}{p_--q_-} \right)^2
- \frac{{\mu_n}^2}{q_-(p_--q_-)} \right]
\label{22}
\end{eqnarray}
with $\mu_1 = 2M$ and  $\mu_2 = 0$. Using the gap equation together with Eq.~(\ref{22}),
one finds two bound states with eigenvalues
$\mu_n^2/(2 p_-)$. 
The particle--anti-particle amplitude $S^{(n)}$ of each eigenstate is given by
\begin{equation}
S^{(n)}(\vec{p},\vec{q}\,) = N^{(n)}(\vec{p}\,)
\frac{\theta(q_-) \theta(p_--q_-)}
{E- \frac{M^2+q_{\perp}^2}{2q_-} - \frac{M^2+q_{\perp}^2}{2(p_--q_-)}}
{\cal V}^{(n)}(\vec{p},\vec{q}\,)
\label{24}
\end{equation}
where $N^{(n)}(\vec{p}\,)$ is a normalization factor.
For the pion in the chiral limit, the amplitude takes a
particularly simple form, 
\begin{equation}
S^{(2)}(\vec{p},\vec{q}\,) = -2{\rm i}N^{(2)}(\vec{p}\,)
\frac{\sigma_{\perp}q_\perp + \sigma_3 M}{q_{\perp}^2 + M^2}
\theta(q_-) \theta(p_- - q_-)\, . 
\label{26}
\end{equation}
This is proportional to the expressions appearing in Eqs. (\ref{dphi1}), (\ref{app1}) and 
(\ref{x6}) in the main text since $p_->0$ and 
\begin{equation}
\left(\sigma_{\perp}q_\perp + \sigma_3 M\right)^2 = q_{\perp}^2+M^2 \ .
\label{app2}
\end{equation}


\newpage

\begin {thebibliography}{99}
\bibitem{Nambu60}
Y. Nambu, Phys. Rev. Lett. {\bf 4}, 380 (1960).
\bibitem{Goldstone61}
J. Goldstone, Nuov. Cim. {\bf 19}, 154 (1961). 
\bibitem{Cash71}
A. Casher, S.H. Noskowicz and L. Susskind, Nucl. Phys. B {\bf 32}, 75 (1971).
\bibitem{Cash73}
A. Casher and L. Susskind, Phys. Lett. B {\bf 44}, 171 (1973).
\bibitem{Cash74}
A. Casher and L. Susskind, Phys. Rev. D {\bf 9}, 436 (1974).
\bibitem{BPP98}
S. J. Brodsky, H.-C. Pauli, and S. S. Pinsky,
Phys. Rep. {\bf 301}, 299 (1998).
\bibitem{GN74}
D. J. Gross and A. Neveu, Phys. Rev. D {\bf 10}, 3235 (1974).
\bibitem{tHooft74}
G. 't~Hooft, Nucl. Phys. B {\bf 75}, 461 (1974).
\bibitem{NJL61}
Y. Nambu and G. Jona-Lasinio, Phys. Rev. {\bf 122}, 345 (1961); {\em ibid.} {\bf 124}, 246 (1961).
\bibitem{Ward50}
J. C. Ward, Phys. Rev. {\bf 78}, 182 (1950).
\bibitem{Takahashi57}
Y. Takahashi, Nuov. Cim. {\bf 6}, 371 (1957).
\bibitem{Miransky93}  V. A. Miransky, {\em Dynamical Symmetry Breaking in Quantum
Field Theory}, World Scientific, Singapore (1993). 
\bibitem{GOR68}
M. Gell-Mann, R. J. Oakes and B. Renner, Phys. Rev. {\bf 175}, 2195 (1968).
\bibitem{R18}
F. Lenz, M. Thies and K. Yazaki, Phys. Rev. D {\bf 63}, 045018 (2001).
\bibitem{R19}
M. Burkardt, F. Lenz and M. Thies, Phys. Rev. D {\bf 65}, 125002 (2002).
\bibitem{LTLY91}
F. Lenz, M. Thies, S. Levit, and K. Yazaki, Ann. Phys. {\bf 208}, 1 (1991).
\bibitem{Schwinger62}
J. S. Schwinger, Phys. Rev. {\bf 128}, 2425 (1962). 
\bibitem{TO93}
M. Thies and K. Ohta, Phys. Rev. D {\bf 48}, 5883 (1993).
\bibitem{Fujikawa79}
K. Fujikawa, Phys. Rev. Lett. {\bf 42}, 1195 (1979). 
\bibitem{BHMY99}
W. Bentz, T. Hama, T. Matsuki, and K. Yazaki, Nucl. Phys. A {\bf 651}, 143 (1999).
\bibitem{Heinzl00}
T. Heinzl,  Nucl. Phys. Proc. Suppl. {\bf 90}, 83 (2000).
\bibitem{IM01}
K. Itakura and S. Maedan, Progr. Theor. Phys. {\bf 105}, 537 (2001).
\end {thebibliography}
\end{document}